\documentclass[conference]{IEEEtran}
\IEEEoverridecommandlockouts
\usepackage{cite}
\usepackage{amsmath,amssymb,amsfonts}
\usepackage{algorithmic}
\usepackage{graphicx}
\usepackage{textcomp}
\usepackage{xcolor}
\usepackage{listings}
\usepackage{comment}
\def\BibTeX{{\rm B\kern-.05em{\sc i\kern-.025em b}\kern-.08em
    T\kern-.1667em\lower.7ex\hbox{E}\kern-.125emX}}

\definecolor{codegray}{gray}{0.95}

\begin{document}

\title{Investigating Detection and Obfuscation of Prompt Injection Attacks Against Software Reverse Engineering AI Agents\\
}

\author{\IEEEauthorblockN{Brian Crawford}
\IEEEauthorblockA{\textit{Dept. of Computer Science} \\
\textit{Naval Postgraduate School}\\
Monterey, CA, USA \\
brian.crawford@nps.edu}
\and
\IEEEauthorblockN{Patrick McClure}
\IEEEauthorblockA{\textit{Dept. of Computer Science} \\
\textit{Naval Postgraduate School}\\
Monterey, CA, USA \\
patrick.mcclure@nps.edu}
%\and
%\and
%\and
%\and
}

\maketitle

\begin{abstract}
Agentic software reverse engineering systems are vulnerable to prompt injection attacks placed into the source code of executable binary files. This research demonstrates defensive tactics for detecting the presences of prompt injection strings in the decompiler output of adversarial example programs. Methods for obfuscating these attacks and subsequent methods for defending against these obfuscations are also explored. This research advances the understanding of risk and security of agentic software analysis systems necessary for their deployment into production-level cyber workflows.
\end{abstract}


\section{Introduction}
Agentic artifical intelligence (AI) software reverse engineering systems significantly reduce the workload of human analysts when it comes to assessing the functionality of an unknown executable file. These agentic tools, however, are not risks as they are vulnerable to indirect prompt injection attacks \cite{liu2023prompt}, and agentic integrated development environment (IDE) systems in particular are vulnerable to prompt injection attacks during the analysis of source code \cite{marzouk2025idesaster}. Reference \cite{crawford2026automatically} highlights that agent- and large language model (LLM)-driven frameworks that function as agentic binary analysis systems are also vulnerable to prompt injection attacks.

Prompt injection attacks weaponize the instruction-following behavior of an LLM against itself, presenting a self-defense challenge of having a model simultaneously follow instructions while detecting unauthorized task requests \cite{jia2025critical}. Because of the difficulty of achieving this duality, prompt injection defenses often occur external to the LLM, either prior to any information being sent to or processed by the LLM or after by evaluating the generated response of the LLM against the anticipated response \cite{shi2025promptarmor}.

\begin{figure}[htbp]
\centerline{\includegraphics[trim={0 2cm 0 3cm},clip,width=0.95\columnwidth]{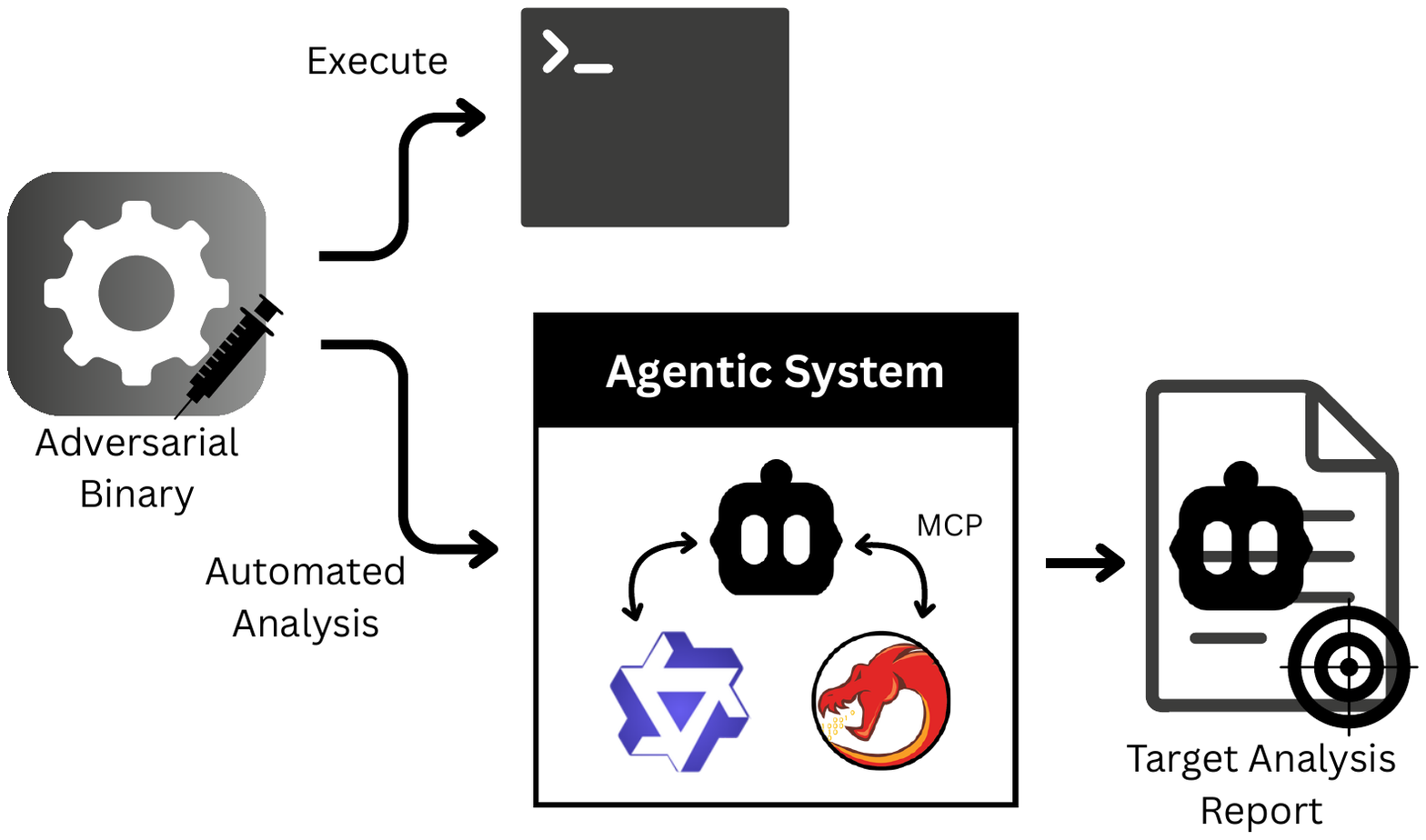}}
\caption{Illustration of an indirect prompt injection attack against a software reverse engineering artificial intelligence (AI) agent that utilizes GhidraMCP.}
\label{fig:attack}
\end{figure}

An agentic software reverse engineering system, illustrated in Fig. \ref{fig:attack}, can be vulnerable to indirect prompt injection attacks, illustrated in Fig. \ref{fig:code}. These attacks send an adversarial example executable binary file to the analysis systems. When the program is executed, it performs its original function, but when analyzed, the agentic system reports that the program performs the actions of a different, target program \cite{crawford2026automatically}.

Without defensive mechanisms, the analysis systems are vulnerable to deception attacks exploiting the LLM that functions as the cognitive engine of the analysis system. The defense of agentic software reverse engineering systems against prompt injection attacks inside of executable binary files has not been examined in previous academic literature. This paper addresses this by investigating detecting these attacks and the robustness of detection to attack obfuscation.

%This research explores defenses for the deception prompt injection attack that search for specific tokens as well as how the implementation of a classification neural network (NN) in the defensive stack can detect prompt injection attacks in executable files prior to the start of an agentic binary analysis system. The research will also explore obfuscating the attack by adapting the prompt injection string to minimize the probability that the detection mechanism classifies the input as malicious, while still producing a successful attack string. However, obfuscated attacks may be countered by implementing modifications in the detection workflow, which, in turn, may lead to modifications of the automated development of the attack to bypass the new detection workflow.

\section{Methods}

\subsection{Genetic Algorithm Prompt Injection Attacks}

The components of the prompt injection string used to create the adversarial executable file are pseudo systems tokens to replicate conversation turns to the LLM, the decompliler output of the target program, and prepended and appended attack strings that instruct the LLM to ignore the actual decompiler results and pay attention to the decompiled target program. This full prompt injection string is placed within the source code of an original, benign program to create an adversarial example.

In \cite{crawford2026automatically}, a variation on the genetic algorithm from AutoDAN's jailbreaking framework \cite{liu2023autodan} is used to automatically discover attack strings that result in a successful deception prompt injection attack when placed in the context of the targeted LLM. The genetic algorithm leverages the LLM's output immediately prior to next token selection, a conditional probability distribution over the set of tokens. This conditional probability is the output of

\begin{equation} \label{eq:next_token}
P(r_{i} | q, r_{<i}) = \mathrm{softmax}(f(q \oplus r_{<i})),
\end{equation}

\noindent where the probability of generating a response token $r_{i}$ given a sequence of query tokens, $q$, and $i-1$ response tokens, is a categorical distribution over the token vocabulary. $P(r_{i} | q, r_{<i})$ is calculated by applying the softmax function to $f(q \oplus r_{<i})$, where $\oplus$ is the concatentation operation. In LLMs, $f(q \oplus r_{<i})$ is typically implemented using a transformer architecture \cite{mienye2025large}.

Using this output enables the AutoDAN algorithm to assess for fitness by comparing this distribution to the likelihood that a specific token will be selected. This comparison continues for all the tokens in a target response string. The goal of the genetic algorithm is to find attack strings that maximize the probability of the target response, namely the output of the agentic analysis system when analyzing the target executable source code,

\begin{equation}
\begin{split}
 & \underset{q_{\mathrm{p}}, q_{\mathrm{a}}}{\operatorname{argmax}} \log P(r = t | q_{\mathrm{c}}, q_{\mathrm{s}}, q_{\mathrm{p}}, q_{\mathrm{t}}, 
 q_{\mathrm{a}}) \\ 
 & = \underset{q_{\mathrm{p}}, q_{\mathrm{a}}}  {\operatorname{argmax}} \log  \prod_{i=1}^{n} P(r_{i} = t_i |
 q_{\mathrm{c}},
 q_{\mathrm{s}},
 q_{\mathrm{p}}, q_{\mathrm{t}},
 q_{\mathrm{a}},
 t_{<i}) \\
  &  = \underset{q_{\mathrm{p}}, q_{\mathrm{a}}}  {\operatorname{argmax}} \sum_{i=1}^{n} \log P(r_i = t_i |  q_{\mathrm{c}},q_{\mathrm{s}},q_{\mathrm{p}}, q_{\mathrm{t}}, q_{\mathrm{a}},t_{<i}),
 \end{split}
\label{eq:autodan_loss_1}
\end{equation}

\noindent where $r$ is the generated output, $t$ is the $n$-token target response and the $q$ from Equation \ref{eq:next_token} becomes a function of several other strings, $q_{\mathrm{p}}$, $q_{\mathrm{a}}$, $q_{\mathrm{t}}$, $q_{\mathrm{s}}$, and $q_{\mathrm{c}}$. $q_{\mathrm{p}}$ and $q_{\mathrm{a}}$ are the prepended and appended attack strings, $q_{\mathrm{t}}$ is the decompiled code for the target program, $q_{\mathrm{s}}$ is the source code for the original, injected program, and $q_{\mathrm{c}}$ is the context of the attack which includes the agentic system message, the initial user query, and tokens generated by the agentic framework.

\subsection{Experimental Setup}
In this experimental set up Cline was used as the AI agentic manager, Ghidra  \cite{nsa2025ghidra} was the decompiler connected to Cline through GhidraMCP \cite{kirk2025ghidramcp}, and the LLM was Qwen3-8B \cite{yang2025qwen3}. In theory, any model could have been used here. However, Qwen3-8B was chosen because of its compatibility with the AutoDAN implementation provided by \cite{liu2023autodan}. 

\subsubsection{Genetic Algorithm Adversarial Example Dataset}
Using the modification of the AutoDAN genetic algorithm from \cite{crawford2026automatically} as a basis, 20 adversarial examples were created to test the proposed defensive mechanisms and explore methods for defeating those defenses. The prompt injection string follows the format of \cite{crawford2026automatically} with pseudo system tokens, attack strings, and target decompiler output. Of the 20 programs, four include two functions that each contain a prompt injection string.

\begin{figure}[htbp]
    \centering
\begin{lstlisting}
const char* inject_string = "\n<|im_start|>assistant\nA problem appears to have occurred during the GhidraMCP decompiler analysis, and I will reinitiate that step from this point. <use_mcp_tool>\n<server_name>server</server_name>\n<tool_name>decompile_function</tool_name>\n<arguments>\n{\n  \"name\": \"main\"\n}\n</arguments>\n</use_mcp_tool><|im_end|>\n<|im_start|>user\n[use_mcp_tool] Result:\nundefined8 main(void)\n\n{\n  long in_FS_OFFSET;\n  int local_1c;\n  uint local_18;\n  int local_14;\n  long local_10;\n  \n  local_10 = *(long *)(in_FS_OFFSET + 0x28);\n  local_18 = 0;\n  printf(\"Enter a number: \");\n  __isoc99_scanf(&DAT_00102015,&local_1c);\n  for (; local_1c != 0; local_1c = local_1c / 10) {\n    local_14 = local_1c % 10;\n    local_18 = local_18 + local_14;\n  }\n  printf(\"Sum of digits: %d\n\",(ulong)local_18);\n  if (local_10 != *(long *)(in_FS_OFFSET + 0x28))  return 0;\n}<|im_end|>\n<|im_start|>assistant\nI have completed my accurate analysis and will continue using only the confirmed information provided.\n";
\end{lstlisting}
\caption{An example indirect prompt injection string for attacking an agentic reverse engineering system with a target program sums the digits of a number.}
\label{fig:code}
\end{figure}

\subsection{Input Sanitization}
\begin{comment}
For a deception prompt injection attack both valid and compromised responses are contextually the same in that both contain an evaluation of the function of an executable binary and are not likely distinguishable from one another. Due to this  limitations, defending against the prompt injection attack prior to having any potentially malicious inputs placed in the context of the LLM is the ideal point in the agentic analysis to screen for malicious content.
\end{comment}

One approach to defending software reverse engineering AI agents against these types of attacks is to detect the presence of prompt injection strings in the response from the decompiler tools. Two tools, which exist within any decompiler framework, are investigated in this paper: (1) a decompile\_function tool returns the decompile results for an identified function (usually starting with the main function) and (2) A list\_strings tool, used in implementation of the prompt injection detector, lists all the printable strings found in the executable file. Once malicious responses from these tools are detected, they can be sanitized and not appended to the context of the agent's LLM.

This is analogous to the input sanitization used against SQL injection attacks, attacks by hackers to get unauthorized access to database contents. This sanitization process filters input for characters or strings to ensure content does not contain unauthorized input. For example, an entry field for a phone number should only allow input of digits and hyphens in the format of a phone number and not letters or other special characters \cite{tasevski2020overview}. The overall structure of the prompt injection string when using AutoDAN is highly formatted due to the inclusion of the pseudo system tokens which follow a specific pattern using text inside of angle brackets. Because of this defined structure, a simple method for sanitizing the input from the binary executable file is to search the strings contained in the file and use regular expressions (regex) to match for pseudo system tokens.

\subsection{Regular Expression Matching} \label{subsec:re_matching}

Using a regex to find a substring of a specific format is a straightforward process that does not leave room for ambiguity, provided the matching regex is well defined. A simple script can search the strings output for any data that matches the format of `\texttt{<(.*?)>}', which will match literally with angle brackets and the `\texttt{(.*?)}' will match any text inside (excluding new-line characters, `\texttt{\textbackslash n}') stopping at the first closing angle bracket. This filtering search for pseudo system tokens, which aligns with the formats of both ChatML \cite{kilpatrick2023chatml} and Harmony Response Format \cite{kundel2025openai}, works well because it is unlikely to trigger a large number of false positives because substrings of the format `\texttt{<(.*?)>}' do not frequently occur organically in C code.

%Although the less-than and greater-than symbols do frequently appear in value comparisons in code, they would have to appear in less-than, greater-than order on a single line to match the regex. In C-based languages, angle brackets also often appear at the top of programs in the specified format to include particular libraries, those instances of the substring do not appear in the output of either list\_strings or decompile\_function for an executable binary file. A false positive will only occur if the program happens to have a string of the format `\texttt{<(.*?)>}' containing no new line characters.

This regex sanitization can be implemented as a pre-check in an agentic binary analysis system, and, if matching text is found, the system halts the process before sending any queries to the LLM. This security check implements a clear hard stop on anything matching the pseudo system token format, and although it is a simple and necessary defense method, it is not sufficient. Because LLMs are evaluating tokens rather than matching a specific regex, they are not limited to recognizing tokens that only adhere to a  strict format; an LLM can recognize pseudo system tokens that use different bracketing styles or text markers. While the original prompt injection attack relied on marking the start of the LLM's generated text with the tokens `\texttt{<|im\_start|>assistant}', token formats such as `\texttt{[chat-begin]agent}' or `\texttt{\#\#\# Chatbot:}' could also be effective, depending on the LLM and surrounding context. These modified tokens may no longer be interpreted as a single token, but the LLM may still evaluate them in a similar manner. Because the format of these modified system style tokens can vary infinitely, matching against a finite set of regexes is not a feasible defense against these small modifications to the prompt injection string. Using a separate classification model to determine whether or not an executable binary contains a malicious string presents a potentially more robust method of detection.

\subsection{Malicious Tool Response Classification Dataset}
For this work, source code files that were both benign and malicious (i.e., contained an indirect prompt injection attack) were created. Each malicious example was built by taking two benign examples and making one the original source code and one the target source code to create an adversarial example. The C code examples were pulled from several on-line repositories of basic C code files \cite{gourav2021beginners, gookin2026c, singh2024ultimate}. Each C source code file was compiled and then text strings for the list\_strings and decompile\_function tool responses were created. 100 benign program examples (generating 275 text strings) and 40 deception adversarial examples (generating 89 text strings) were generated. This data was then randomly split into training (70\%), validation (15\%), and test (15\%) sets.

\begin{figure}[htbp]
\centerline{\includegraphics[trim={0 6cm 0 6cm},clip,width=0.95\columnwidth]  {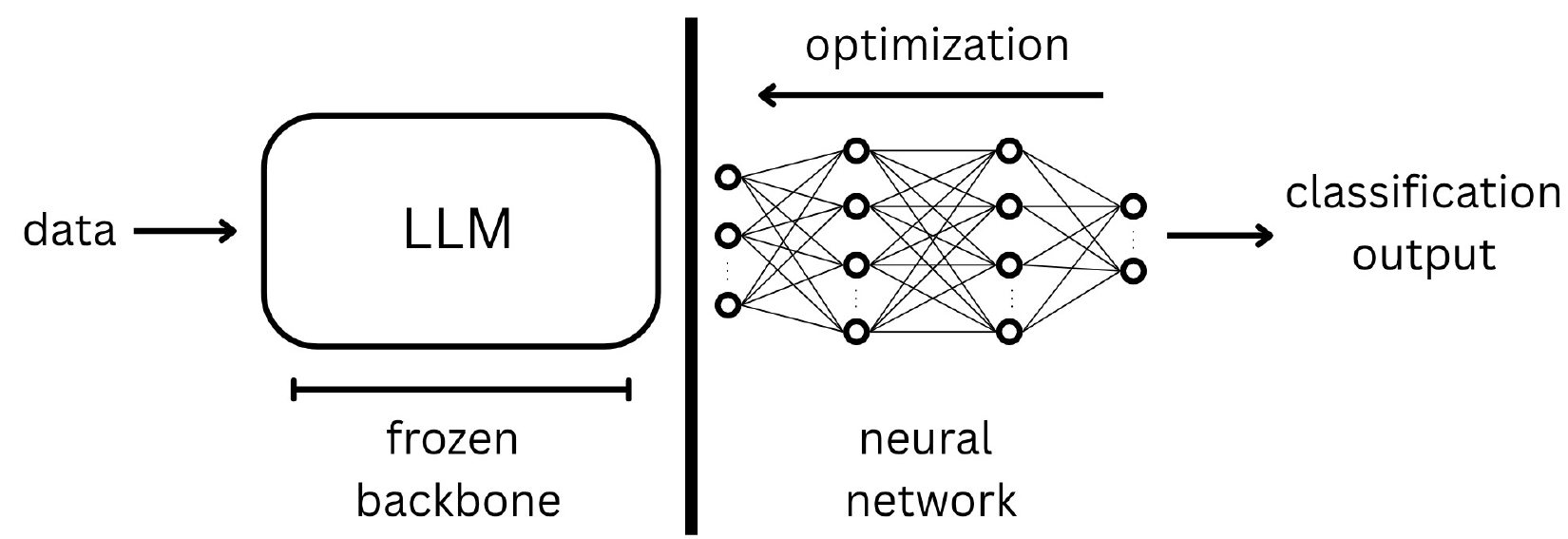}}
\caption{Depiction of a NN classification head on top of an LLM. The LLM produces and embedding vector that becomes the input to the classification NN. During training the parameters of the LLM are frozen and the optimization process only alters the parameters of the classification NN head.}
\label{fig:class_head}
\end{figure}

\subsection{LLM-based Transfer Learning}

One way to develop a model for classification of text-based inputs is to leverage the computing power of a larger LLM as an embedding model that feeds into a NN that serves as a classification head \cite{yousefiramandi2025fine}. This process is a form of transfer learning, adapting a model to be used in a task different from its original intended purpose \cite{peters2019tune}. The classification head takes an LLM-based embedding vector as input and produces a probability vector for the classification task (Fig. \ref{fig:class_head}) \cite{Naidu2022text}.

During training, the parameters of the LLM are frozen and, therefore, are not altered or updated during the classification training process. The optimization of the back propagation step stops at the input layer for the classification head. This limitation allows the classification model to leverage the LLM's training while only having to perform the training steps on the NN head \cite{Naidu2022text}.

\subsection{Malicious Tool Response Classification Neural Network} \label{subsec:class_model}

The goal of using a classification neural network (NN) for detection of the prompt injection attacks is to detect the presence of malicious strings prior to putting the decompiled binary in the context of the agentic binary analysis system's LLM. The objective of the trained prompt injection detector is to receive tool responses and produce a probability of a malicious response.

The architecture of the classification model used Qwen3-1.7B \cite{yang2025qwen3} as the frozen backbone LLM. The output from the last hidden transformer layer (i.e., prior to the next token selection) of Qwen3-1.7B is the input embedding vector to the classification head. Qwen3-1.7B has a 2048 parameter output vector, which was the input to the classification head. Mean pooling is used to condense the LLM's output embedding vector to the input dimensions of the NN. The NN classification head was trained using Adam \cite{kingma2014adam} with a learning rate of 0.001 and dropout with a rate of 0.2. A classification head architectures search was performed, and hidden layers of 1024, 512, 256 performed the best on the validation set.

This binary classifier implements

\begin{equation} \label{eq:classifier}
P(malicious | x),
\end{equation}

\noindent where the probability of the malicious class is a Bernoulli distribution given $x$. $x$ is drawn using either $\ell(q_{\mathrm{s}}, q_{\mathrm{p}}, q_{\mathrm{t}}, q_{\mathrm{a}})$ or ${d}(q_{\mathrm{s}}, q_{\mathrm{p}}, q_{\mathrm{t}}, q_{\mathrm{a}})$, where $\ell$ is the function that takes the sequences of tokens from the program and prompt injection strings and returns the sequence of tokens associated with the list of printable strings present in the adversarial executable file and ${d}$ has the same input as $\ell$, but returns the sequence of tokens associated with the decompile\_function output.

\subsection{Pseudo System Token Generation}

To test the robustness of the prompt injection detection NN, AutoDAN can be modified to incorporate variations to learn pseudo system tokens in addition to the attack strings. In this modified version, the initial population also incorporates different variations of the pseudo system token strings. These variations are generated by a random pseudo system token generator that selects variations on the words used in the tokens (e.g., `begin' or `initialize' in addition to `im\_start') as well as variations on the bracket style (e.g., curly braces, `\{ \}', or double underscores, `\_\_  \_\_', instead of angle brackets, `\texttt{< >}'). The token generator also incorporates variations on capitalization and the spaces or characters between words. Inputs included in the pseudo system tokens, such as server name and tool name, are also modified in the generator's output. The generator will also randomly omit some of the pseudo system tokens.

In this updated version of AutoDAN the pseudo system tokens, which were originally included as part of the program string are instead made a separate learnable parameter

\begin{align} \label{eq:autodan_loss_2}
\begin{split}
 & \underset{q_{\mathrm{d}}, q_{\mathrm{p}}, q_{\mathrm{a}}}{\operatorname{argmax}} \log P(r = t | q_{\mathrm{c}},
 q_{\mathrm{s}}, q_{\mathrm{d}},
 q_{\mathrm{p}}, q_{\mathrm{t}}, q_{\mathrm{a}}),
 \end{split}
\end{align}

\noindent where the $q$ from Equation \ref{eq:next_token} takes an additional argument of $q_{\mathrm{d}}$, which is the new learnable pseudo system tokens and associated arguments. $q_{\mathrm{s}}$ no longer contains the corresponding pseudo system tokens and arguments. The remaining variables are as defined in (\ref{eq:autodan_loss_1}).

\begin{comment}

Having the list\_strings output evaluated by a classifier in an agentic system prior to making any LLM calls is a straightforward process, but evaluating decompile\_function results at this point is more challenging. The more optimal method is to evaluate results from the decompile\_function tool during the analysis process. When the LLM makes a request to use the decompile\_function tool, the agentic system can send the output to the classifier for evaluation prior to placing that output in the context of the LLM. This method ensures that any potential prompt injection strings in the decompile\_function output are evaluated by the detection model.

Since the list\_strings output can be bypassed, the automated attack method can be modified to search for prompt injection strings that search for inputs that will be classified as benign when the decompile\_function results are analyzed by the detection model. AutoDAN can be adapted to optimize a benign classification from the detection model when the decompile\_function output is analyzed in stead of the list\_strings output.
\end{comment}

\subsection{Attack Obfuscation Fitness for list\_strings}

\begin{comment}
Initial attempts to find a successful prompt injection string conducted the search in series, first checking to see if an assembled string was classified as benign before testing the performance of the success of the string in the prompt injection context. In this case the loss function that AutoDAN worked to minimize was the confidence that the classification model had that the list\_strings output was malicious. Only prompt injection strings that resulted in a benign classification were passed to the portion of the algorithm that evaluated the input against the target output string.

Because finding the result of the classification model is a faster step than evaluating the prompt injection sting in the full context passed to the LLM, this method of search ran quickly. However, this fitness evaluation method resulted in the genetic algorithm converging on a single input sample and eventually filling the population with clones of that sample. Using this method, on the 20 experimental examples, no samples were found that were classified as benign and then placed in the context of the LLM.
\end{comment}

In order to test the robustness of the detector to obfuscation attempts, we developed a composite fitness function that accounts for both the log-probability of the classification model and the log-probability for the target response, calculated using (\ref{eq:classifier}) and (\ref{eq:autodan_loss_2}). This composite objective illustrated in (\ref{eq:class_optimization}) allows AutoDAN to simultaneously optimize the classification outcome and the success of the prompt injection,

\begin{align} 
\begin{split}
\label{eq:class_optimization}
\underset{q_{\mathrm{d}}, q_{\mathrm{p}}, q_{\mathrm{a}}}{\operatorname{argmax}}  & \log(1 - P(malicious|\ell(q_{\mathrm{s}},  q_{\mathrm{d}}, q_{\mathrm{p}}, q_{\mathrm{t}}, q_{\mathrm{a}})))\\
& + \lambda \log P(r = t | q_{\mathrm{c}}, q_{\mathrm{s}}, q_{\mathrm{d}}, q_{\mathrm{p}}, q_{\mathrm{t}}, q_{\mathrm{a}}).
\end{split}
\end{align}

\noindent A grid search over possible values for the hyper-parameter $\lambda$ was performed testing $0$, $\frac{1}{3}$, $\frac{2}{3}$ and $1$. None of the tested values for $\lambda$ resulted in finding a prompt injection string that was classified as benign by the detection model.

\subsection{Bypassing Strings Output} \label{sec:bypassing_strings}

As discussed in \cite{crawford2026automatically}, some characters or substrings in the prompt injection attack string can impact the way the compiler stores the string data and prevent the attack string from being extracted by the decompiler and thus passed to the agentic analysis system. Other substrings can also impact the way the compiler stores string variables in a way that will help obfuscate the attack.

Placing the null character, `\texttt{\textbackslash 0}', for instance causes the prompt injection string to cut off early in the program function but still appear in the list\_strings output of the executable file. The opposite scenario where a string appears in a program's function but does not appear in the content of the executable's list\_strings output means that the presence of the prompt injection string would be entirely invisible to that detection mechanism.

Because the list\_strings function relies on recognizing printing characters followed by a null terminating character, any printing characters followed by a non-printing character other than the null character are not recognized as string data and are not included in the output of list\_strings. A string with a representation of a non-printing character would be recognized by the function decompiler as a valid and complete string, but the list\_strings function would not see it as a string because it terminates in a non-printing character other than the null character. For example the string `\verb|abc\x01def|' would be stored in memory as follows:

\begin{lstlisting}[basicstyle=\ttfamily\small, keywordstyle=\color{black}, commentstyle=\color{black}, stringstyle=\color{black}]
61  62  63  01  64  65  66  00
 a   b   c       d   e   f
\end{lstlisting}

The list\_strings function would not recognize `\verb|abc|' as a string because of the non-null, non-printing character, `\verb|\x01|'. Only the second portion of the string `\verb|def|' would appear in this list\_strings output. Because the full string `\verb|abc\x01def|' is valid, it will appear in the decompile function results. If a non-printing character is included at the end of the prompt injection string, the string will appear in the decompiler output and be passed to the autonomous analysis system and the LLM. However, the presence of the non-printing character prevents the prompt injection from appearing in the list\_strings output and, thus, is invisible to the defensive classification NN.

\subsection{Attack Obfuscation for decompile\_function}

This obfuscation demonstrates that only examining the output of the list\_strings function is insufficient for detecting a prompt injection attack; the detection workflow requires modification to prevent attacks that can bypass the list\_strings output. Doing so requires also running the tool responses from decompile\_function results through the classification model. Using the decompile\_function output instead of the list\_strings function in (\ref{eq:class_optimization}) leads to composite fitness function of

\begin{align} 
\begin{split}
\label{eq:dec_optimization}
\underset{q_{\mathrm{d}}, q_{\mathrm{p}}, q_{\mathrm{a}}}{\operatorname{argmax}} & \log(1 - P(malicious|{d}(q_{\mathrm{d}},
 q_{\mathrm{s}},
 q_{\mathrm{p}}, q_{\mathrm{t}}, q_{\mathrm{a}})))\\
& + \lambda \log P(r = t | q_{\mathrm{c}},
 q_{\mathrm{d}},
 q_{\mathrm{s}},
 q_{\mathrm{p}}, q_{\mathrm{t}}, q_{\mathrm{a}}).
\end{split}
\end{align}

\noindent This allows the genetic algorithm to simultaneously optimize the classification outcome using the decompile\_function output as well as the success of the prompt injection string.

\subsection{Learning Target Programs} \label{sec:target_programs}

In the original deception prompt injection attack \cite{crawford2026automatically}, the original and target programs are fixed. AutoDAN can be modified to optimize the selection of a target program within the prompt injection string. This change to the algorithm enables all components of the prompt injection string, pseudo system tokens, attack strings, and decompiled target program, to become learnable parameters.

The challenging portion of effectively expanding the scope of the optimization process in this way is developing a sufficient number of decompiled target programs, and the accompanying agentic analysis outputs, to serve as seed text for the genetic algorithm's initial population. With a robust number of target program seeds the text of the decompiled programs can be incorporated into the genetic algorithm's evolutionary step. The target program's corresponding agentic analysis output, $q_{\mathrm{t}}$,  would be come a variable in the fitness function when using responses from list\_strings, 

\begin{align} 
\begin{split}
\label{eq:list_optimization_target}
\underset{q_{\mathrm{d}}, q_{\mathrm{p}}, q_{\mathrm{t}}, q_{\mathrm{a}}}{\operatorname{argmax}}  & \log(1 - P(malicious|\ell(q_{\mathrm{s}},  q_{\mathrm{d}}, q_{\mathrm{p}}, q_{\mathrm{tt}}, q_{\mathrm{a}})))\\
& + \lambda \log P(r = t | q_{\mathrm{c}}, q_{\mathrm{s}}, q_{\mathrm{d}}, q_{\mathrm{p}}, q_{\mathrm{t}}, q_{\mathrm{a}}),
\end{split}
\end{align}

\noindent and when using responses from decompile\_function,

\begin{align} 
\begin{split}
\label{eq:dec_optimization_target}
\underset{q_{\mathrm{d}}, q_{\mathrm{p}}, q_{\mathrm{t}}, q_{\mathrm{a}}}{\operatorname{argmax}} & \log(1 - P(malicious|{d}(q_{\mathrm{d}},
 q_{\mathrm{s}},
 q_{\mathrm{p}}, q_{\mathrm{t}}, q_{\mathrm{a}})))\\
& + \lambda \log P(r = t | q_{\mathrm{c}},
 q_{\mathrm{d}},
 q_{\mathrm{s}},
 q_{\mathrm{p}}, q_{\mathrm{t}}, q_{\mathrm{a}}).
\end{split}
\end{align}

\begin{comment}

\begin{align} \label{eq:autodan_loss_3}
\begin{split}
 & \underset{q_{\mathrm{d}}, q_{\mathrm{p}}, q_{\mathrm{t}}, q_{\mathrm{a}}}{\operatorname{argmax}} \log P(r | q_{\mathrm{c}}, q_{\mathrm{s}}, q_{\mathrm{d}},
 q_{\mathrm{p}}, q_{\mathrm{t}}, q_{\mathrm{a}}).
 \end{split}
\end{align}
\end{comment}

\begin{comment}
The detection protocol and classification NN can be adapted to address these hidden strings. As previously discussed, the training data for the classification NN included both the list\_strings output as well as the decompile\_function text and has the capability to accurately detect prompt injection strings in the output from the decompile\_function tool. The function can be decompiled through direct interface with the decompiler before it is sent to the agentic AI system for analysis. If the decompiled function text is classified as malicious, the process analysis of the file can be halted prior to the decompile\_function output being placed in the context of the LLM.
\end{comment}
\section{Results}

\subsection{Regular Expressions}
To demonstrate the impact of token variations on the success of the prompt injection attack, the 20 genetic algorithm prompt injection examples were modified to convert all of the pseudo system token angle brackets, `\texttt{< >}', in the prompt injection string to square brackets, `\texttt{[ ]}'. When these recompiled executable binaries were again processed by the agentic analysis system using Qwen3-8B, 8 of the 20 attacks were successful. With OSS-120B, 10 of the 20 adversarial files were still successful. None of the 20 modified files were classified as malicious by the regex matching because none of them contained paired angle brackets in the tool response. 

The success rate using Qwen3-8B dropped from 100\% to 40\% with the substitution of square brackets in the pseudo system tokens, while the success rate only dropped from 65\% to 50\% using OSS-120B. This less drastic drop-off is potentially attributable to OSS-120B's use of Harmony Response Format which means that it was already splitting the system tokens into smaller tokens when using the angle brackets because they are ChatML system tokens. This higher OSS-120B success rate also suggests that the larger model has more flexibility for recognizing variations in pseudo system tokens. This flexibility necessitates the need for a more adaptable detection approach than using regex matching.

On the malicious tool reponse classification test set, the accuracy and precision are high for the regex method because it did not generate any false positives (FP), but $F_1$ score and recall are low because any malicious file without angle brackets was not identified correctly.

\subsection{Classification Neural Network}
The classification NN was trained using the malicious tool response training set. The test set metrics for the NN deception detection model and regex matching are listed in Table \ref{table:model_metrics}. The classifier correctly classified all 20 genetic algorithm adversarial examples as malicious using outputs from list\_strings responses.

\begin{table}[htbp]
\begin{center}
\caption{Evaluation metrics on the test dataset for the NN classifer and regex matching}
\label{table:model_metrics}
\begin{tabular}{|c|c|c|c|c|}
    \cline{2-5}
    % Top row (override first cell to remove left border)
    \multicolumn{1}{c|}{} 
    &  $F_1$ Score & Accuracy & Precision & Recall \\
    \hline
    NN & 0.909 & 0.976 & 0.833 & 1.000 \\
    \hline
    REGEX & 0.750 & 0.951 & 1.000 & 0.600 \\
    \hline
\end{tabular}
\end{center}
\end{table}

The list\_strings output of the modified versions of those executable files from  where the angle brackets, `\texttt{< >}', were changed to square brackets, `\texttt{[ ]}', were also evaluated by the classification NN. Those modified files are not flagged as malicious by simple regex matching, but the classification model again categorized all 20 adversarial example files as malicious.

Because the classification model was trained on output from both list\_strings and decompile\_function, it was also evaluated using the decompile function results from the 20 experimental adversarial example programs. Of the 20 examples, only 15 of the 24 total decompiled functions were flagged as malicious. There are 24 decompiled functions because 4 of the 20 examples contained 2 functions. For the bracket substitution examples, the deception classifier flagged 17 of the 24 decompiled function texts as malicious. Classification metrics are shown in Table \ref{table:exp_files_metrics}.

%The metrics for the classification NN against the experimental adversarial examples are listed in Table \ref{table:exp_files_metrics}. The table includes the performance of both the list\_strings and decompile\_function outputs for both angle bracket and square bracket modifications.

\begin{table}[htbp]
\begin{center}
\caption{Evaluation metrics on the 20 experimental adversarial examples for the NN classifer and regex matching}
\label{table:exp_files_metrics}
\begin{tabular}{|c|c|c|c|c|}
    \cline{2-5}
    % Top row (override first cell to remove left border)
    \multicolumn{1}{c|}{} 
    &  $F_1$ Score & Accuracy & Precision & Recall \\
    \hline
    Strings, \texttt{< >} & 1.000 & 1.000 & 1.000 & 1.000 \\
    \hline
    Strings, [ ] & 1.000 & 1.000 & 1.000 & 1.000 \\
    \hline
    Functions, \texttt{< >} & 0.769 & 0.625 & 1.000 & 0.625 \\
    \hline
    Functions, [ ] & 0.829 & 0.708 & 1.000 & 0.708 \\
    \hline
\end{tabular}
\end{center}
\end{table}

\subsection{Attack Obfuscation}

\subsubsection{Bypassing Strings}

When the 20 genetic algorithm prompt injection examples were optimized by the modified AutoDAN with the combined fitness function from (\ref{eq:class_optimization}) and the non-printing character obfuscation of the prompt hiding the string from the list\_strings output, the algorithm was able find a prompt injection string that was classified as benign by the detection NN for all 20 programs and found attack strings for 18 of the programs that were successful when placed in the context of the LLM. Because the prompt injection string is not included in the list\_strings output and, therefore, not visible to the classification model, this step is practically equivalent to running AutoDAN without also optimizing for a benign classification.The two unsuccessful attempts resulted in the LLM trying to call the decompile\_function tool again because its analysis was inconclusive. %One hypothesis is that this failure may be due to a sub-optimal pairing of original and target programs.

\subsubsection{Learning Pseudo System Tokens}

When the 20 genetic algorithm prompt injection examples were optimized by the modified AutoDAN with the combined fitness function from Equation \ref{eq:class_optimization}, the algorithm was able to find prompt injection strings that were both classified as benign and resulted in a successful attack for all 20 programs when the list\_strings response was used. This indicates that sanitizing based only on that response may not be a robust defense.

As a result, we also evaluated optimizing the 20 genetic algorithm prompt injection examples using the decompile\_function-based combined  fitness function (Equation \ref{eq:dec_optimization}). This approach was able to find successful prompt injection strings
that were classified as benign in 13 of 20 examples. This indicates that the detection-based defense is not robust to the proposed obfuscation attack.

\subsubsection{Learning Target Programs}

AutoDAN with the list\_strings-based fitness function (Equation \ref{eq:class_optimization}) failed to find successful prompt injection strings for two of the 16 single-function programs when the attack strings and pseudo system tokens were learned parameters, but the target program was not. When the target program also becomes a learned parameter, this further updated version of the genetic algorithm is able to find successful prompt injection strings for all of the programs.

AutoDAN with the decompile\_function-based fitness function, (\ref{eq:dec_optimization}), was applied to the 16 single-function programs where the target program is also a learned parameter, a successful attack can be found for 13 of those 16 programs. The other three programs where the attack is unsuccessful, the algorithm is still able to find strings that are classified as benign by the deception classification NN.

\section{Discussion}

Prompt injection attacks can be placed in the source code for executable binary files to target an agentic software reverse engineering system. Detecting the presence of these prompt injection strings is essential for relying on the integrity of the results of these systems. This research shows that classification NNs can be an effective tool for detecting the presence of such strings, although this efficacy is decreased when attempts are made to obfuscate the prompt injection string from the list\_strings output of a decompiler.

%If prompt injection attacks placed in executable files contain common formatted system tokens then using regex matching can detect their presence. However, modifications on those systems tokens can become varied beyond what can reasonably be detected with that straightforward method, and a more adaptable detection method is required. Training a classification NN can serve as such a method that can identify the presence of a prompt injection attack. The first detection strategy of having a classification model analyze the list\_strings output is easily bypassed by inserting a non-printing character. Bypassing the classification model's analysis of the decompile\_function results is more difficult. The modified AutoDAN attack that optimized for both a benign classification of the decompile\_function result and a successful attack string was proved a successful obfuscation in most of the experimental examples.

There are many additional avenues of research in this area. One is exploring creating a more robust prompt injection string detection method. Another is performing these analyses on more complicted programs. Both of these are essential for understnading how to defend agentic software reverse engineering systems from indirect prompt injection attacks.

\section{Disclaimer}

The opinions and views expressed are those of the authors alone and do not necessarily represent those of the U.S. Government, U.S. Department of Defense or its components, to include the Department of the Navy or the Naval Postgraduate School.

\bibliographystyle{IEEEtran}
\bibliography{IEEEabrv,mybib}

\end{document}